# Digital Electronics for the Pierre Auger Observatory AMIGA Muon Counters

O. Wainberg,[a,b,*] A. Almela,[a,b] M. Platino,[a] F. Sanchez,[a] F. Suarez,[b] A. Lucero,[a,b] M. Videla,[a] B. Wundheiler,[a] D. Melo,[a] M. Hampel,[a] and A. Etchegoyen[a,b]

[a] *Instituto de Tecnologías en Detección y Astropartículas (CNEA, CONICET, UNSAM),*
*Centro Atómico Constituyentes, San Martin, Buenos Aires, Argentina*

[b] *Universidad Tecnológica Nacional, Facultad Regional Buenos Aires,*
*Buenos Aires, Argentina*

*E-mail*: oscar.wainberg@iteda.cnea.gov.ar

ABSTRACT: The "Auger Muons and Infill for the Ground Array" (AMIGA) project provides direct muon counting capacity to the Pierre Auger Observatory and extends its energy detection range down to 0.3 EeV. It currently consists of 61 detector pairs (a Cherenkov surface detector and a buried muon counter) distributed over a 23.5 km$^2$ area on a 750 m triangular grid. Each counter relies on segmented scintillator modules storing a logical train of '0's and '1's on each scintillator segment at a given time slot. Muon counter data is sampled and stored at 320 MHz allowing both the detection of single photoelectrons and the implementation of an offline trigger designed to mitigate multi-pixel PMT crosstalk and dark rate undesired effects. Acquisition is carried out by the digital electronics built around a low power Cyclone III FPGA. This paper presents the digital electronics design, internal and external synchronization schemes, hardware tests, and first results from the Observatory.

---

[*] Corresponding author.

# Contents

# 1. Introduction

The Pierre Auger Observatory detects the highest known cosmic ray energies in the universe [1] starting from ~1 EeV[1]. The Observatory spans over 3000 km$^2$ of pampas at 1400 m above sea level in the Malargüe and San Rafael departments in the province of Mendoza, Argentina. It works as a hybrid system; a Surface Detector (SD) and a Fluorescence Detector (FD) [2]. The former consists of over 1600 water-Cherenkov detectors deployed on a 1500 m triangular grid while the latter consists of four groups of air fluorescence telescopes (a group with nine and the remaining three with six FD telescopes) overlooking the SD array. The SD samples the lateral density distribution of the secondary particles. It has an almost 100% duty cycle[2]. The FD works only on clear moonless nights and ,while active, is sensitive to the elements. Consequently it has an average duty cycle of ~13%. By measuring the fluorescence light emitted by atmospheric $N_2$ during shower development, it provides near calorimetric measurements of particle numbers as showers develop.

The "Auger Muons and Infill for the Ground Array" [3] (AMIGA) has a denser array than the SD, covering 23.5 km$^2$. It is located about 6 km east from the Coihueco FD building. It comprises pairs of surface stations and Muon Counters (MC). The surface stations measure both the muonic and electromagnetic components of high energy Extensive Air Showers (EAS) whilst the MCs measure the muonic component.

Following the introduction section, the surface setup is presented with a glance at surface to underground synchronization. After a brief description of the underground scintillator module, the underground electronics hardware is described. The section on the digital-board FPGA code describes the main digital electronics modules implemented in the FPGA, core of the MC electronics. The FPGA front end, responsible of sampling the 64 channels at 320 MHz, is particularly explained in detail. The system synchronization responsible of data coherence is presented afterwards. Finally, hardware and software validation tests are presented, from initial laboratory setups to Observatory data.

## 1.1 Physics requirements on the electronics

As charged cosmic rays acquire higher energies, stronger acceleration mechanisms are required. At some stage, it is suspected that our galactic sources reach a maximum limit. Therefore, the highest energy nuclei, arriving to earth, should come from further away in the universe. Might that be the case, at a certain energy, a change should occur in the cosmic rays composition, the flux or both. By increasing the detector-grid density, AMIGA, with its 750 m infill area plus a programmed denser area, targets the energy range from 0.05 to 10 EeV where the transition is expected to take place. AMIGA's muon counters take a direct measurement of the shower muonic component, consequently contributing with an independent entity to composition analysis. Its electronics is designed to accurately count muons over a range of muon numbers to precisely measure a muon lateral distribution function with small systematic timing uncertainty.

The physics requirements on the electronics stem for the need to count individual muons and maintain synchronization with the surface stations that measure the complete air-shower signal. The first is achieved by a direct muon counting approach that avoids the need for deconvolution of an integrated signal. The last is accomplished by the synchronization of the muon counter to the nearby surface station on the low-level trigger and the synchronization to the surface array on the higher-level trigger.

---

[1] 1 EeV = $10^{18}$ eV.
[2] Duty cycle is only limited by technical availability.

Counting of muons implies a one-bit electronics for either the presence or absence of a muon and a strategy to distinguish muons and discard cross talk. This was attained by requiring the electronics to work at the mean SPE width thus directly counting Single photoelectron (SPE). Recently, a signal-integrator board of all 64 strips has been implemented with an analog to digital converter (ADC) for those MCs which happen to lie close to the core where signal pile-up occurs.

AMIGA scintillator modules are divided into discrete scintillator bars. Discriminators, located in the analog boards, convert the incoming signal to either a '0' for undetected values, or a '1' for a signal above a predefined threshold, with no intermediate values. The resulting continuous digital signal is sampled and acquired by the FPGA, located in the digital board, in real-time at 320 MHz, or 3.125 ns between samples. Selected sampling interval, in the range of the width of a SPE at 30% of its peak voltage, makes possible individual SPE detection.

Muon counter event acquisition is synchronized at the lowest (hardware) level to the surface stations through a dedicated triggering line originated in the surface stations. MC electronics maintains synchronization, through a time tagging scheme, mostly implemented in the digital board FPGA. An event data request trigger, received by the surface radio, is sent from the surface to the underground microcontroller through an Ethernet line. The digital board FPGA searches for the requested event and retrieves data. An offline counting strategy searches for the muon traces by inspecting the individual SPE signatures. As the vast majority of contaminating events produce only a sole SPE, by requiring at least a '1', 'X'[3], '1' for a muon footprint, accidentals including a possible double sample due to a wider than average SPE, are removed. Consequently, most of the accidental data, such as crosstalk or thermal photoelectrons is discarded.

## 1.2 AMIGA brief description

At present, all 750 m infill Surface Stations are operative. A 750 m hexagon called "Unitary Cell" is being installed with two long 10 m$^2$ plus two short 5 m$^2$ muon detector scintillator modules, yielding a 30 m$^2$ detection surface at each of the vertices and its center. The 10 m$^2$ scintillator modules are made of 64 scintillator strips 41 mm wide, 10 mm high and 4 m long. The 5 m$^2$ modules use shorter 2 m long strips. Two additional 30 m$^2$ MCs placed at the hexagon center and right vertex positions form two 30 m$^2$ + 30 m$^2$ pairs called twins. These 60 m$^2$ detectors are intended to experimentally verify data consistency between their two constituents. To date, five vertices and their center are fully operative with a 10 m$^2$ detector each and, in addition, a twin has four 10 m$^2$ plus four 5 m$^2$ scintillator modules, totaling 120 m$^2$ of detection surface. Figure 1 shows the installed MC distribution in the Observatory.

---

[3] An 'X' can be either a '0' or a '1'.

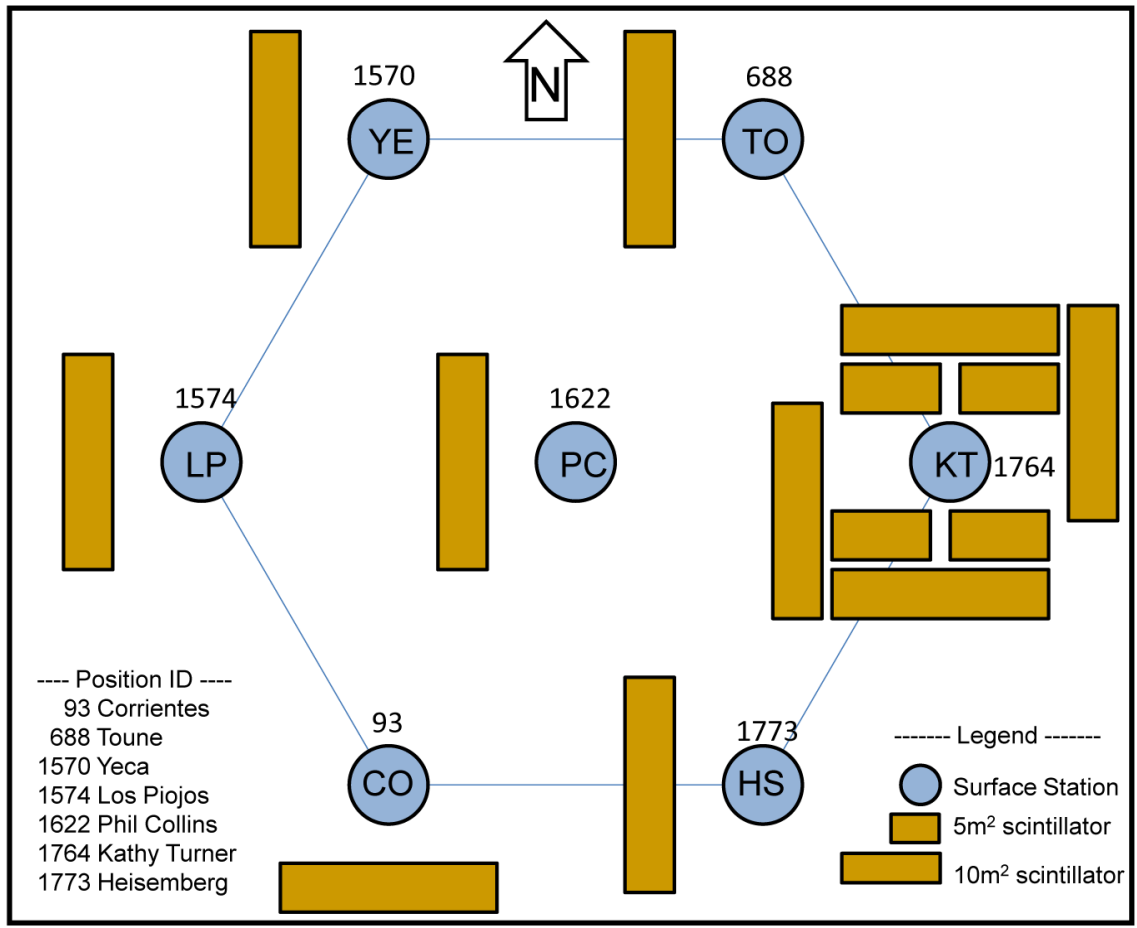

**Figure 1:** AMIGA unitary cell working detectors distribution[4]. Light blue circles represent surface stations. Orange short and long rectangles represent 5 m$^2$ and 10 m$^2$ detectors respectively. Diagram not to scale.

As counters work in the open field, conventional shielding based on lead is not practical either from the logistical or from the ecological point of view. Instead, soil from the installation site is used as electromagnetic shielding. Chemical and density studies at different locations of the AMIGA site yielded a homogenous density of (2.38 ± 0.05) g/cm$^3$ to 3 m depth [4] thus requiring a 2.25 m soil column for shielding of 535 g/cm$^2$.

Each buried scintillator module houses an underground electronics setup. It receives photons originated in the scintillator bars, acquires data in real-time, storing EAS data in a local memory. The use of multiple modules requires a distributed electronics approach. On a request from the central facility, data are retrieved. Surface electronics located above the local surface station electronics receives event data from the associated underground electronics modules. A Wi-Fi radio transmits data to a concentrator linked to the Central Data Acquisition System (CDAS). A solar panel setup together with charger and batteries delivers the required power to the electronics. Figure 2 shows the electronics set mounted at the center of a scintillator module in the laboratory. Identified in the figure are the optical connector, the PMT, and all the electronics boards.

---

[4] As for 01/12/2013.

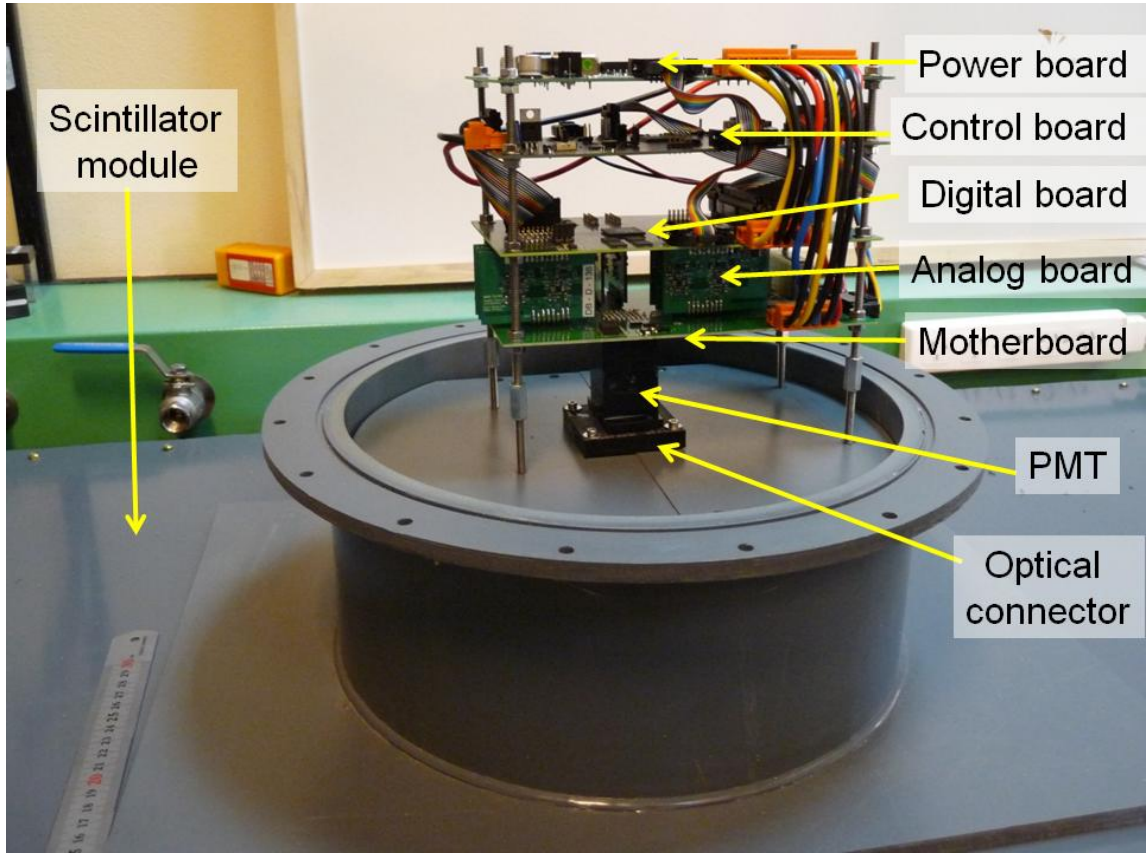

**Figure 2:** Underground electronics setup mounted on top of a laboratory scintillator module.

**1.3 Surface detector**

SD electronics is built around an electronics setup called the Local Station (LS) [5] with an acquisition data rate of 40 MHz. It works under a hierarchical trigger system [6]. T1 is the lowest level, which solely depends on the individual detector. On a trigger occurrence, detector data are locally stored with an associated timestamp. A T2 trigger promotes the T1s that will be used by CDAS to decide which event data will be permanently stored. All T2 timestamps are transmitted to CDAS. On T2 receptions, CDAS looks for groups of detectors forming a compact pattern compatible with EAS physics. On detection of a group, a T3 trigger is broadcasted to the detectors within the triggering zone requesting the locally stored associated data. In response, participating surface stations send their data to CDAS. Higher level offline triggers complete the selection process.

**1.4 Data storage**

Muon counters acquire data at 320 MHz. Every 3.125 ns a new 64 bit word is acquired. Each bit stores the digitized value associated with a corresponding scintillator bar. An event spreads over 6.4 µs comprising 2048 samples[5]. Each new event is uniquely identified by an associated timestamp. The T1 time bin is used to synchronize a surface station to a muon counter event by identifying a predefined position in the acquired data frame, corresponding to the same absolute time in both systems. In the surface station it corresponds to the 256$^{th}$ bin while in the MC it is preset by a dedicated register. Data from the first bin up to the T1 position corresponds to pre-trigger samples while data from T1 to the last position correspond to post-trigger samples. 2048 events are stored locally in a logical ring buffer on a RAM bank. With an average storage rate of 100 events per second, an event can be recalled up to approximately 20 seconds after its occurrence. Data for recalled T3 events are permanently stored by CDAS, unrecalled events are overwritten.

## 2. Muon counter surface software and hardware

MC surface electronics comprise a solar energy power system, surface-station trigger extraction and distribution, a microcontroller board, a Wi-Fi wireless radio and an Ethernet switch. Two ARTESA AP-170 170 W solar panels, connected to a 24 V 20 A regulator, charge a 24 V battery bank made of two sealed deep cycle 12 V 150 Ah series connected batteries. Power is distributed to the underground electronics under a Power over Ethernet (PoE) arrangement that shares a single twisted-pairs cable bundled with the transmitted data (instead of using a separate power cable).

Local Station hardware is built around a Cyclone (I) EP1C12Q240I7 FPGA (or two ACEX EP1K100QI208-2 chips in older versions) and a PowerPC 403GCX embedded microcontroller running an OS9000 operating system. Muon counters are synchronized to the surface station LS internal FPGA T1 trigger. A code added to the LS FPGA and microcontroller generates triggering and tagging signals for the underground electronics and surface MC microcontroller board. Figure 3 shows the scheme for surface synchronization with the underground electronics and radio. On a T1 occurrence in the LS FPGA (1), a local timestamp is generated (2) and a stream indicating the triggering point and an associated timestamp called Local Time Stamp (LTS) is sent from the FPGA to the underground electronics through a Low-Voltage Differential Signaling[6] (LVDS) driver (3). This timestamp is only recognized in the MCs. After an event ending (4), the microcontroller is interrupted and a second timestamp called the GPS[7] timestamp (GTS), used by the surface station, is generated, identifying the associated surface station event (5). The GTS is sent from the microcontroller to the FPGA (6). Timestamps from surface-station T2-promoted events are sent to CDAS. The event's previous LTS is associated to the GTS and the pair is sent, with LVDS hardware levels, to the MC microcontroller (7). Inside the microcontroller, a 2048 entry table, the same as for locally stored underground events, stores a translation table between the two linked timestamps. For a potential event of interest, CDAS broadcasts a T3 data recall to all nearby triggered surface stations. On T3 reception by the surface-station radio, the MC surface microcontroller recovers the required timestamp from the radio stream (8). The received GTS is searched for in the microcontroller's table. On a match, the linked LTS is fetched (9) and a T3 data recovery command, with the requested LTS, is broadcasted to all MC associated underground electronics through the ethernet link (10).

---

[5] Early installed electronics setups store half this number of samples.
[6] LVDS is defined in the TIA/EIA-644 standard.
[7] Global Positioning System.

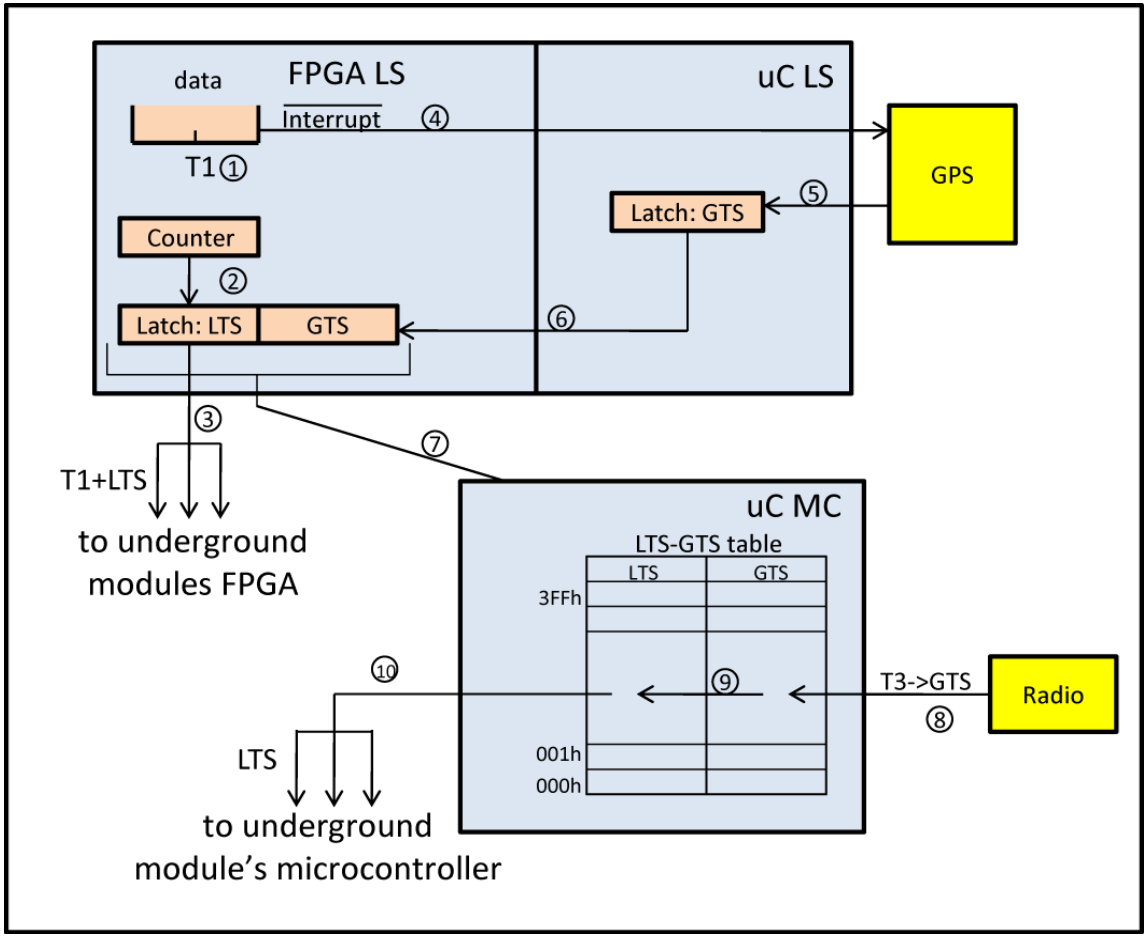

**Figure 3:** Surface electronics synchronization scheme. From event acquisition (1) to (7) to data request (8) to (10).

Figure 4 shows a frame of the dedicated T1 line. The rising edge at (1) indicates the T1 bin. After 200 ns at '1' (2) and 300 ns at '0' (3), a 24 bit LTS timestamp is transmitted, with the most significant bit (MSB) first, at a 100 ns per bin rate. The frame spans 2.9 µs (4), less than the 6.4 µs of a MC event.

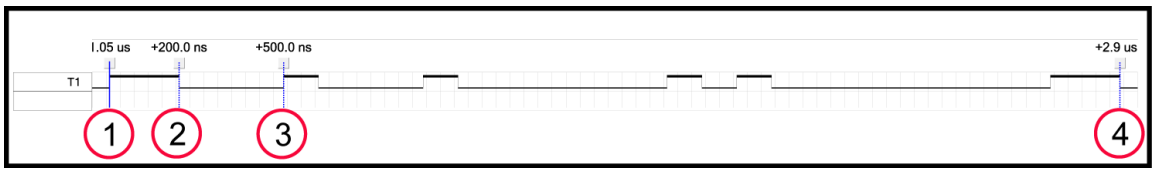

**Figure 4:** T1 trigger pulse rising edge (1) synchronize surface station events with underground muon counter events. Following data stream tags the triggered events with a timestamp. This stream is transmitted in real-time on a dedicated line. Positions (1) to (3) show T1 bin synchronization, (3) to (4) show the 24 bit timestamp data tag.

An R52n-M radio from Mikrotik (working under the IEEE 802.11n standard in the 2.4 GHz band with a maximum power output of 23 dBm, coupled to a 20 dBi parabolic antenna) communicates with the concentrator located on top of the Coihueco hill near the FD building. The concentrator links to CDAS via a point to point backbone connection. A nine port Ethernet switch RB493 from Mikrotik interconnects surface to underground electronics through category 5e twisted pair cables. Ethernet usage allows the interconnection of up to 253 underground modules to a single Surface Station[8].

## 3. Underground module

Each of the scintillator modules contains 64 scintillator bars [7] grouped as two sets of 32 bars placed side by side. Extruded bars are made of Dow Styron 663W polystyrene doped with 1% PPO and 0.03% POPOP by weight and coextruded with $TiO_2$, acting as a photon reflector, coating all the bar except for a central groove. A wave length shifting fiber (WLS) BCF-99-29AMC is glued inside the scintillator groove with optical cement BC-600, both from Saint Gobain. On top of the glued fiber, an aluminum foil tape acts as a photon reflector. Impinging muons on scintillator bars induce photon emission in the region of 410 nm. Some of the emitted photons arrive at the fiber, whose excited molecules decay, emitting photons isotropically with a mean wavelength of 485 nm. Photons propagating within the total internal reflection solid angle eventually reach the ends of the fibers. On one side, the 64 WLS fibers end up in an optical connector coupled to a 64 pixel PMT. On the other side, fiber endings are finished in black matt paint to avoid reflection[9].

## 4. Underground electronics

Underground electronics is responsible for data acquisition in real time and retrieval of data on request. It is composed of a set of boards stacked together.

### 4.1 Motherboard

The motherboard contains a high quantum efficiency Ultra Bi-Alkali (UBA) 64 pixel H8804-200MOD PMT from Hamamatsu [8][9] arranged in an eight by eight matrix. Each of the 2 mm by 2 mm down looking PMT photocathodes faces a WLS fiber end on the mirror-polished face of the optical connector. A negative high voltage (HV) module C4900-01 was chosen to polarize the PMT. A negative photocathode arrangement has the advantage of not biasing the anode's small signal with the high voltage supply. This allows a direct current (DC) coupling to the amplifiers with no overvoltage protection.

The H8804 PMT can operate, within datasheet specifications, to -1000 V. As a reduction in PMT gain is expected due to ageing over the years, the high voltage is set at -950 V, close to the maximum admissible value but with a reasonable reserve for future compensation. The HV is controlled through a close loop. A 12 bit digital to analog converter (DAC) sets the desired voltage, while a HV divider connected to an ADC monitors the actual HV. An overvoltage clipping circuit built around a TL431 programmable reference protects the PMT from an excessive high voltage[10], such as might occur due to an unexpected software malfunction, by limiting the high voltage to -1040 V.

---

[8] Provided power supply capacity meets power consumption requirements.
[9] Even though reflected photons increase the light yield, the extra transit time overcomes the benefit for this application.

**4.2 Analog boards**

Analog boards convert PMT anode pulses to a continuous stream of digital signals that are fed to the FPGA in the digital board. As single photoelectron[11] mean values relationship between the 64 pixels of the same PMT [8] can differ by up to a factor of $2^{12}$, gain differences should be equalized. Gain correction by adjustment of the amplifiers' feedback resistors would not only complicate board assembly because of different resistor values required for particular pixels, but would also affect board replacement due to these personalized values. Moreover, since relative pixel gains may vary over time, a fixed corrective scheme might fail in the long term. Thus, digitization is performed by adjustable-threshold discriminators comparing the instantaneous pixel voltage level against a reference as shown in Figure 5.

The output of a comparator[13] is in the active state when the voltage on the positive input is higher than the voltage on the negative input, regardless of the common voltage[14]. On the other hand, a higher voltage on the negative input brings the output to the inactive state. Figure 6 shows that, for a given pulse shape, the output of the discriminator will be the same for different pulse amplitudes as long as the discriminator thresholds to peak pulse voltages relationship is kept constant (i.e. $V_{th1}/V_{p1} = V_{th2}/V_{p2} = V_{thn}/V_{pn}$). As a result, the output of a discriminator with an adjustable threshold (and variable input signal gain) will be undistinguishable from the same discriminator fed by an adjustable-gain amplifier (and a fixed threshold).

---

[10] The C4900 HV supply can reach -1250 V.
[11] In this paper SPE is always used as a measure of peak voltage magnitude, not as a charge.
[12] AMIGA PMTs are selected for a maximum relative gain between pixels of 2.
[13] Comparators are discriminators in this application.
[14] Within device working specifications.

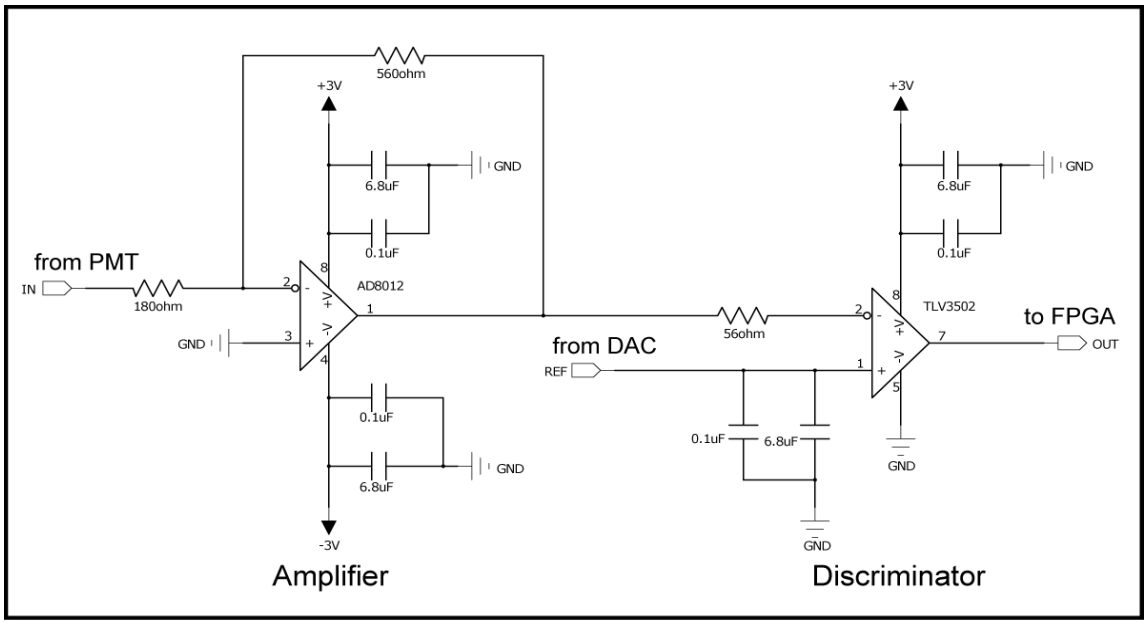

**Figure 5:** One analog board channel composed of an amplifier and a discriminator. (Serial octuple DAC, that sets the discriminator reference, is omitted for simplicity).

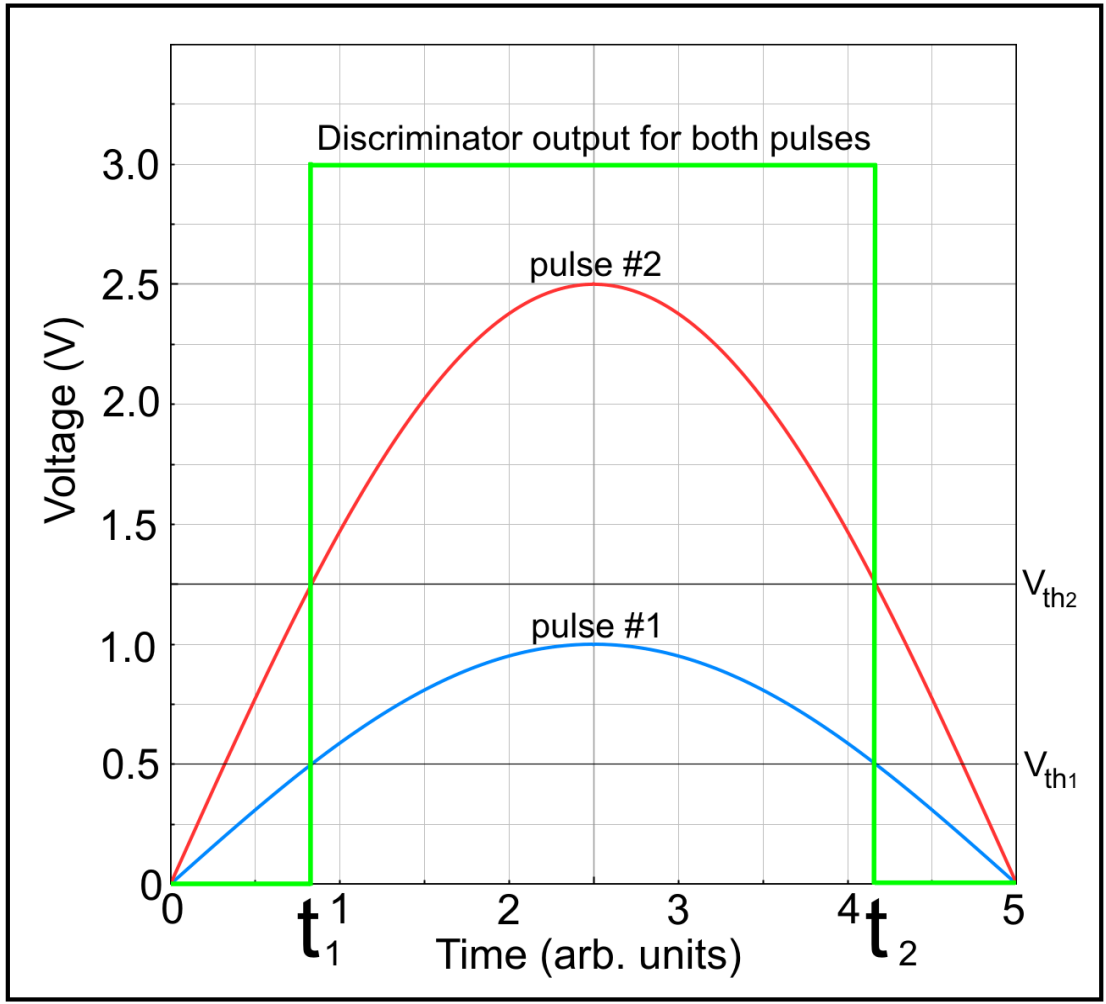

**Figure 6:** The (simulated) discriminator response, in green, is the same for different pulse amplitudes with equal threshold to peak voltage relationships. Given that $V_{th1} / V_{pulse1\_peak} = (0.5 / 1.0 = 50\%) = V_{th2} / V_{pulse2\_peak} = (1.25 / 2.5 = 50\%)$ at $t_1$ both pulses cross their respective threshold references. In blue: $V_{pulse1}(t_1) = V_{th1}$. In red: $V_{pulse2}(t_1) = V_{th2}$. The same, with the opposite direction, occurs during $t_2$: $V_{pulse1}(t_2) = V_{th1}$ and $V_{pulse2}(t_2) = V_{th2}$. As a result, the discriminator responds to both pulses with an identical output.

A 12 bit octuple serial low-power DAC TLV5630 sets individual thresholds for each of the analog board channels. Threshold discriminating values are set around 30% of a <SPE> [10] at the amplifier output. When a PMT anode voltage exceeds the pixel threshold, the discriminator output enters an active state. Output signals are one bit continuous 3.3 V CMOS compatible levels.

### 4.3 Digital board

The digital board receives the 64 discriminated pixel signals from the analog boards and the trigger dedicated line from the surface electronics. It acquires real time events in response to a T1 request. On a T3 request, acquired event data are sent to CDAS via the control board, surface electronics and radio. The digital board is built around a Cyclone III EP3C25F324I7 [11] RAM-

– 11 –

based FPGA. At power on, the FPGA code is sent by radio and uploaded via a standard Joint Test Action Group (JTAG[15]) interface from the control board. This scheme allows a remote code upgrade at any time.

The EP3C25F324I7 controls a 8 MWord by 32 bit memory bank composed of four static RAM Cypress CY62187EV30 [12] chips. A low 2% working duty cycle alongside very low power memory chips yields an average power consumption of only 6 mW for the whole bank. The clock is provided by a 50 ppm 40 MHz local oscillator. The FPGA internal PLL generates 80 MHz and 320 MHz internal clocks. The 64 outputs from the analog boards' discriminators are directly read by the FPGA. Discriminator references, provided by the DACs in each of the eight analog boards, are programmed from the FPGA through a buffered Serial Peripheral Interface (SPI) bus. The FPGA Input-Output (IO) port voltage is provided by a 3.3 V Low Drop-Out (LDO) regulator; core voltage is provided by a 1.2 V LDO regulator. The 10 layer digital board with the FPGA and static RAM memories with FBGA packages is shown in Figure 7.

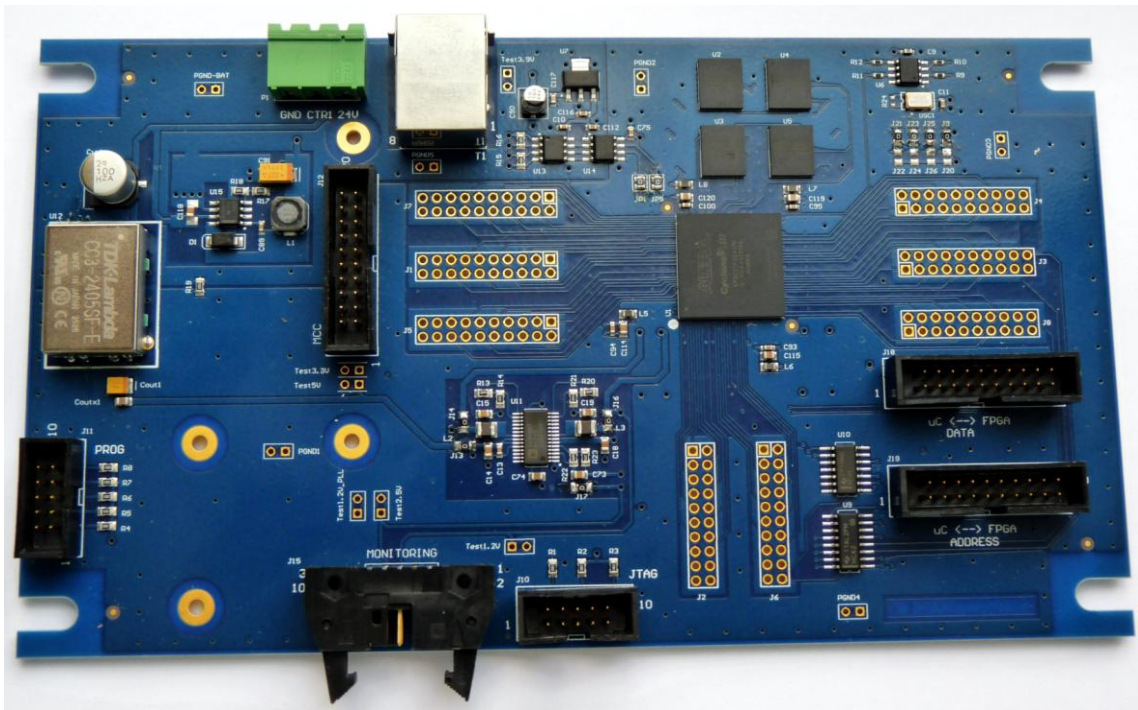

**Figure 7:** Digital Board built around a Cyclone III FPGA in a 10 layer board.

**4.4 Control board**

The control board is built around a NXP LPC2468 [13] microcontroller with an ARM7TDMI-S 32 bit core. Under uClinux [14], it is responsible for communication with surface electronics through a 100 Mbit per second Ethernet link. An uBoot bootloader [15] initializes software from the LAN. It not only allows an easy remote upgrade but, as the bootloader is locally stored in non-volatile memory, bugged software will not lock the system, forcing an on-site reboot. The control board monitors vital parameters such as working voltages and temperature, sending the information to a monitoring server. Monitoring data are used in real time to trigger alarms and, if a system crash should occur, should also help in determining the causes. Similar to the microcontroller code, the FPGA binary code is received remotely through the radio. This code is programmed using a standard JTAG hardware interface with adapted JRunner software from Altera.

---
[15] IEEE 1149.1 Standard Test Access Port and Boundary-Scan Architecture.

### 4.5 Power board

Underground power is provided from the surface batteries through power over Ethernet. As the cable length is of the order of 25 m, depending on the physical conditions of the terrain, a common ground reference point for the whole distributed system cannot be implemented. Consequently, the power board isolates incoming power from the underground electronics with a RECOM RP30-4812SEW DC-DC converter. Switching regulators produce the different voltages used by the analog and digital boards with high efficiency. LDO regulators on local boards reduce noise with minimal power loss.

## 5. The digital-board FPGA code

An important design challenge was to achieve a SPE detection capability, on a statistical basis, with a low cost FPGA[16]. As muon counters and surface stations data are associated, their electronics main clocks were set to work with their frequencies related by a power of two ($2^n$) factor. This relationship simplifies their raw data correlation. Therefore and since the surface station is clocked with a 40 MHz source, an acquisition frequency of $2^3$ * 40 MHz = 320 MHz was chosen. Further to this, the sampling interval of 1 / (320 MHz) = 3.125 ns is close to the mean SPE width, so the electronics is able to count SPEs. Moreover, 320 MHz is within the maximum allowed working frequency. Even though some FPGA parts can work with frequencies over 400 MHz, since interconnectivity resources are limited, the workable speed decreases as resource usage increases. Furthermore, some FPGA parts such as the memory blocks (like an M9K in a Cyclone III -I7 device) have maximum working frequency lower than 320 MHz. As a result, it would not have been possible to work directly with a 320 MHz clock. Working at a lower speed allows full usage of the available resources with a good timing safe margin. Consequently, a working frequency of 80 MHz was chosen for all the FPGA circuitry but the front end that is clocked with a 320 MHz source.

### 5.1 Front end

In order to reduce high speed interconnection resources whilst operating most of the FPGA with an 80 MHz clock, a very simple, yet efficient, front end is implemented. Each of the 64 outputs originating in the discriminators, that are located in the analog boards, are sent to the FPGA. The FPGA samples the inputs at 320 MHz and converts each individual line to four simultaneous lines running under a single 80 MHz clock. As a result the 64 inputs, sampled at 320 MHz, are transformed into 64 * 4 = 256 lines with a slower 80 MHz actualization rate. Figure 8 shows the schematic for one bit. All parts except the counter are independently implemented for each of the 64 channels. The top D flip-flops form a serial-in/parallel-out four bit shift register clocked with the 320 MHz clock. After four or more clock cycles the parallel output stores a four bin history spanning over (1/320 MHz) * 4 = 12.5 ns = (1/80 MHz). A two bit binary counter clocked with the 320 MHz clock has a $2^2$ * (1/320 MHz) = 12.5 ns repetition rate, the same as the shift register history span. Q2 decoded output is set to '1' each time the counter cycles through the '10' sequence. This signal enables the four bit latch formed by the middle flip flops and clocked with the 320 MHz source. After each activation of Q2, every four 320 MHz clock cycles, the latch copies the shift register output. As a result, the latch is updated at an 80 MHz rate. The lower latch, clocked by the 80 MHz clock in phase with the 320 MHz clock, completes synchronization to the 80 MHz domain. Further data acquisition is carried out through a 256 bit wide bus with the slower clock.

---

[16] Around 300 FPGA integrated circuits will be used in the underground electronics of the complete MC array.

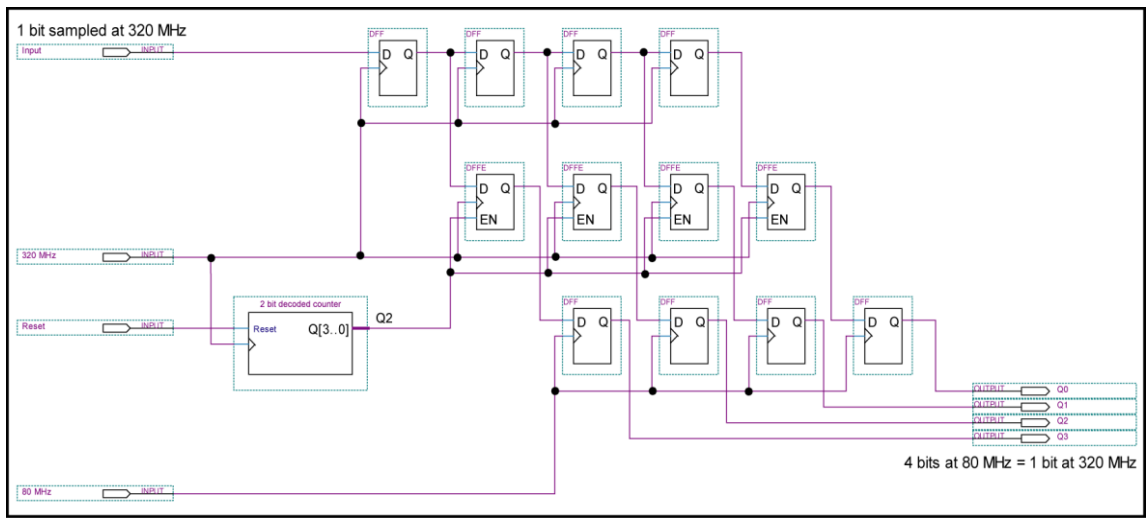

**Figure 8:** Conversion from one line sampled at 320 MHz to four lines at 80 MHz in the FPGA front end. Q2 from the two bit binary decoded counter loads four sampled bits every four 320 MHz clock cycles.

### 5.2 Event acquisition

An event is composed of 2048 words 64 bit wide. Each word represents the state of the 64 pixels. These can be in inactive or active state indicating possible photon detection. Since data are acquired on the nanosecond scale, starting event acquisition with T1 is not a feasible option for the reason that when acquisition is initiated, an important part of the desired event has already occurred. To avoid data loss, acquisition is carried out on a continuous basis whether a trigger occurs or not. A circular buffer scheme as described in a simplified form in Figure 9 and Figure 10 is implemented. At any moment, it holds the past acquired 2048 bin record corresponding to 2048 bin * 3.125 ns/bin = 6.4 μs of data. While waiting for a T1, the pre-T1 counter is disabled and data acquisition is active. On T1 arrival (1) $\overline{Q}$ of the $SR$ flip flop (see Figure 9) passes to the inactive state (2) releasing the counter from the forced load. Consequently, the counter counts downwards a predefined number of bins, 1536 by default (3). After zero is reached, carry activation (4) stops acquisition. At that moment, the circular buffer contains 1536 words post-T1 and 2048 - 1536 = 512 words pre-T1, holding event data prior to T1 reception by the underground electronics. On filling, a second identical (not shown) circular buffer takes over from the first to avoid a dead time. The frozen pre-T1 counter holds the first

– 14 –

event position: $MOD_{2048}$ (512+1536) = 0 (5). A latched address alongside a new event signal (6) finishes event acquisition.

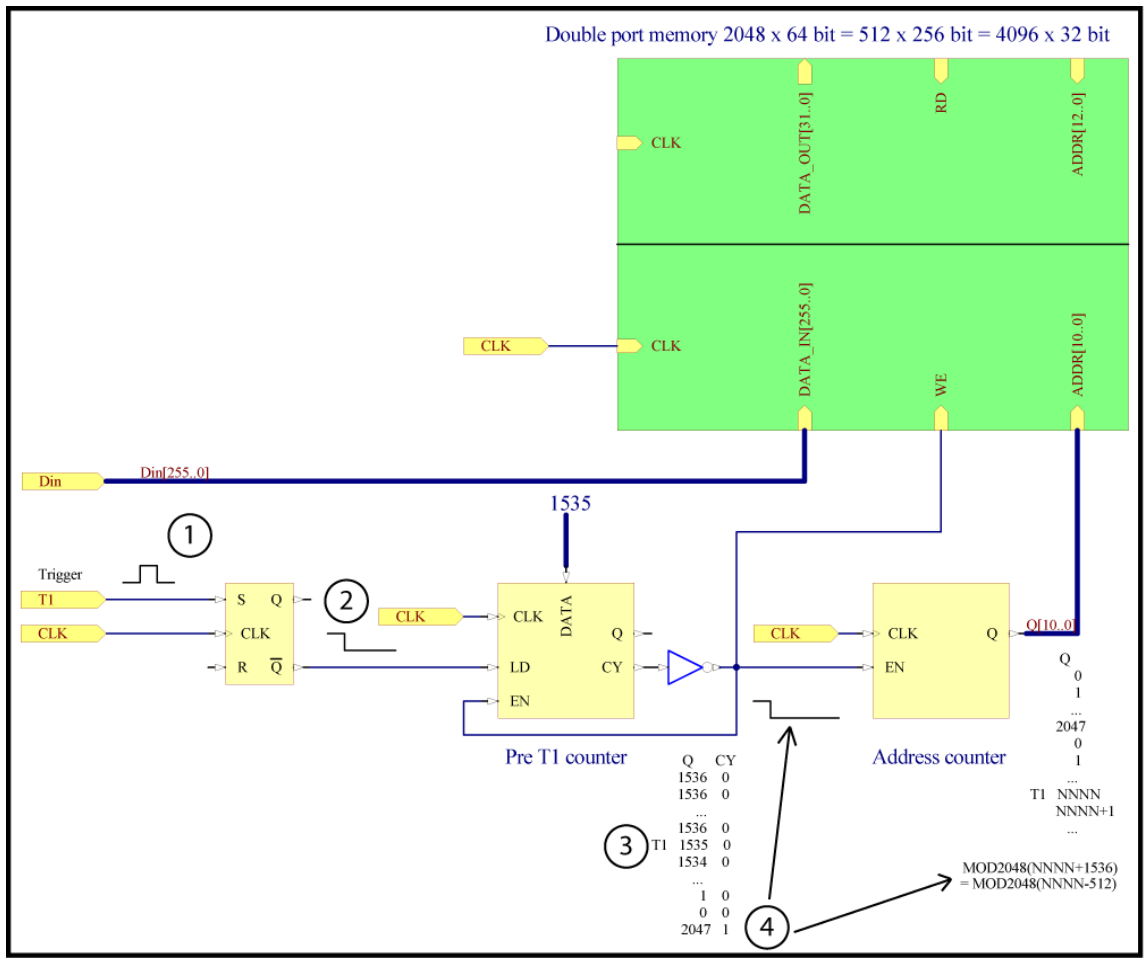

**Figure 9:** Event acquisition in FPGA part I. Circular buffer acquisition stops a predefined (1536 by default) number of cycles after T1 trigger.

Both circular buffers are implemented within a single double port RAM block. This approach permits independent read and write circuitries with different data port widths.

Independent ports allow a very simple hardware construction: the input (write) port is connected to the front end outputs while the output (read) port is directed to the external RAM data bus. Different port widths permit accommodation of the 256 bit input bus to the 32 bit external RAM data bus. The MSBs on both ports are used to select one of the two circular buffers on the fly. A single circuitry is used for both logical input buffer ports. Output port circuitry is implemented in the same fashion.

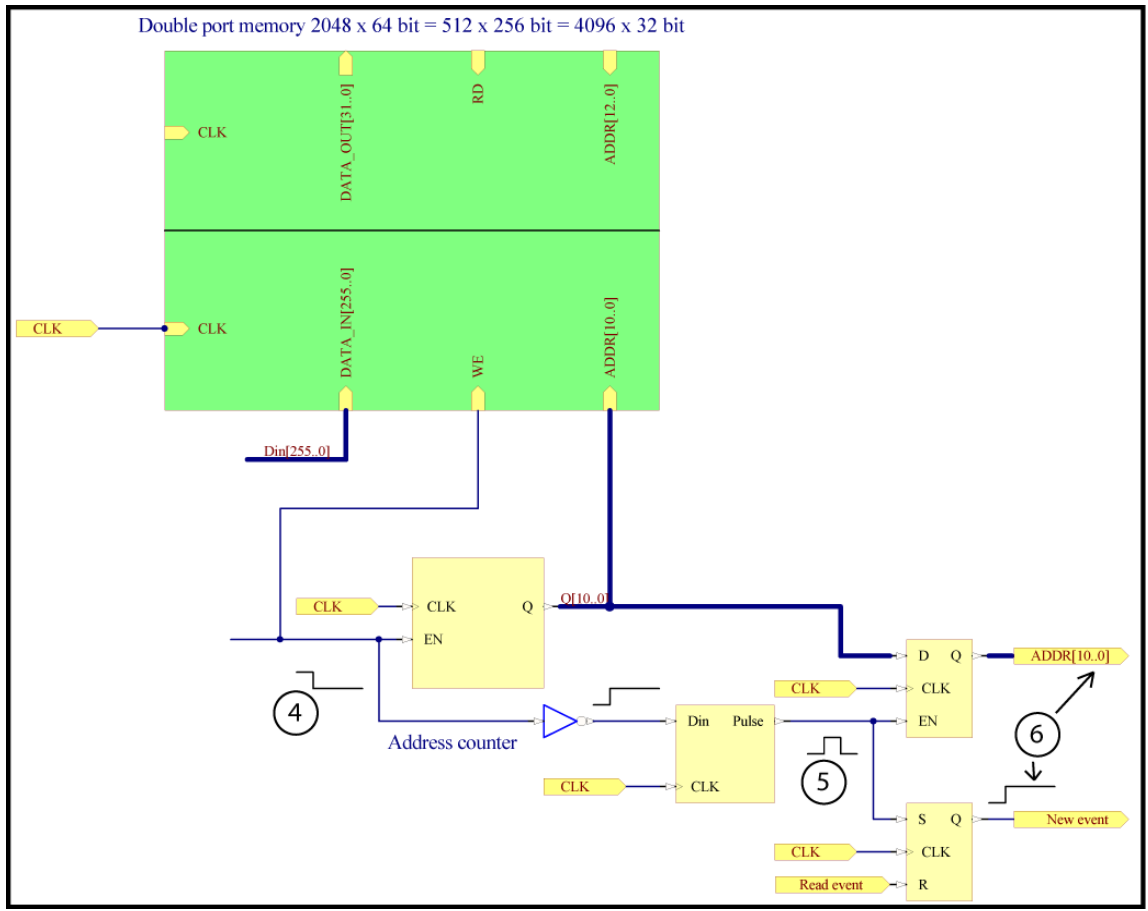

**Figure 10:** Event acquisition in FPGA part II. After an event acquisition, triggered by T1, the address of the first acquired event word alongside a new-event signal is generated for further data storage in the external RAM bank.

### 5.3 External memory management

An external RAM memory bank stores 2048 events in a logical circular buffer. The memory bank is configured as an 8 MWord space 32 bit wide. It uses a standard asynchronous protocol consisting of a plain 23 bit address space, a 32 bit data bus, a $\overline{RD}$ signal for reading, $\overline{WR}$ signal for writing and a $\overline{CE}$ signal to enable a memory transaction. Memory access can be initiated by three sources; it can be written from one of the two T1 event acquisition buffers or read from the T3 buffer. The FPGA implements direct memory access (DMA) type transference with 75 ns access time per word totaling 307.2 µs per event. This cycle time is 20 ns longer than the CY62187EV30 minimum allowed time, yielding a good timing margin yet the total event write time is less than the associated surface station time.

### 5.4 T1 trigger

A dedicated line transmits T1 and an associated timestamp from the surface to underground electronics in real time. The rising edge of a 200 ns pulse indicates the triggering point and initiates event acquisition. 300 ns later, a serialized timestamp is sent at a 10 MHz rate. As a new event cannot occur before the end of the previous one, there is no possibility for a timestamp to be misinterpreted as a false trigger. A trigger pulse stripped from the signal triggers an event acquisition. The associated timestamp is stored alongside the event data.

### 5.5 T3 trigger

Acquired events are stored in the RAM bank with an associated timestamp table. On T3 assertion, the timestamp table is searched for the received LTS. On a match, the position points to the data in the external memory. In the rare case of a not found tag, a flag indicates the unavailability of the requested information. Requested event data are copied to an internal buffer and the microcontroller is notified about the new event availability.

### 5.6 DAC control

Each of the eight analog boards has an eight channel DAC [16]. DACs are serially programmed by an SPI interface. Data received through the DIN input is clocked on the falling edge of the SCLK clock. FS input acts as a chip enable. The first two signals are common to all the DACs while the last is independent for each board. As a daisy chain is avoided, a malfunctioning DAC will not disable others. The FPGA takes the burden of programming up to 80 registers in a row from the microcontroller. A table with a set of 80 twelve bit registers, 64 for each of the discriminator threshold levels plus 16 control registers, holds the values to be programmed. On finishing, all DACs outputs are updated simultaneously through a common, to all boards, $\overline{LDAC}$[17] signal that transfers programmed values to external pins.

### 5.7 Microcontroller interface

The FPGA communicates with the control board through a set of registers seen by the microcontroller as an external memory bank composed of 32768 words, 16 bit wide. Only active registers are implemented, leaving space for future add-ons. Most of FPGA functions are executed by programming the appropriate data registers followed by the activation of the corresponding execution bit. When appropriate, successful (or unsuccessful) finalization is acknowledged by status bits. Functions such as T3 data recovery return their data in the memory bank. A few registers are used for debugging.

### 5.8 Calibration

Different pixels have different <SPE> values. Discriminator thresholds are used to compensate for this. If a channel has a higher than desired gain, its <SPE> value might produce a larger number of active FPGA samples. On the other hand, a lower gain would reduce the number of active samples. Setting a higher discriminator threshold value through the corresponding DAC channel reduces the number of active samples, compensating the effect of the high gain. Conversely, a lower threshold value is used to compensate a lower channel gain.

64 thirty-two bit registers, one for each scintillator segment, are continuously incremented for each FPGA positive sample[18] of the corresponding channel. As a result, each register accumulates the number of positive samples from power on. An additional 64 bit register counts time clocked by the 80 MHz source. On command, all registers (accumulators and time) are latched at once. By subtracting two readings and dividing the result by the time difference, an average rate is obtained for each channel. The difference between a channel average and its associated set-point is a measure of the deviation from the desired working point. This deviation is used to calculate a new channel threshold for the corresponding DAC. Following the DAC output update, the new rate moves towards the desired set-point. Figure 11 shows the closed loop auto-calibration adopted method.

---

[17] Load DAC. In a TLV5630 [16] DAC, when this signal is low the outputs copy the state of the programmed value through the SPI interface.
[18] In the installed code the rate depends on the background muon rate as well as PMT dark rate. On a future upgrade, muons might be counted instead of active samples according to the adopted counting strategy [17] .

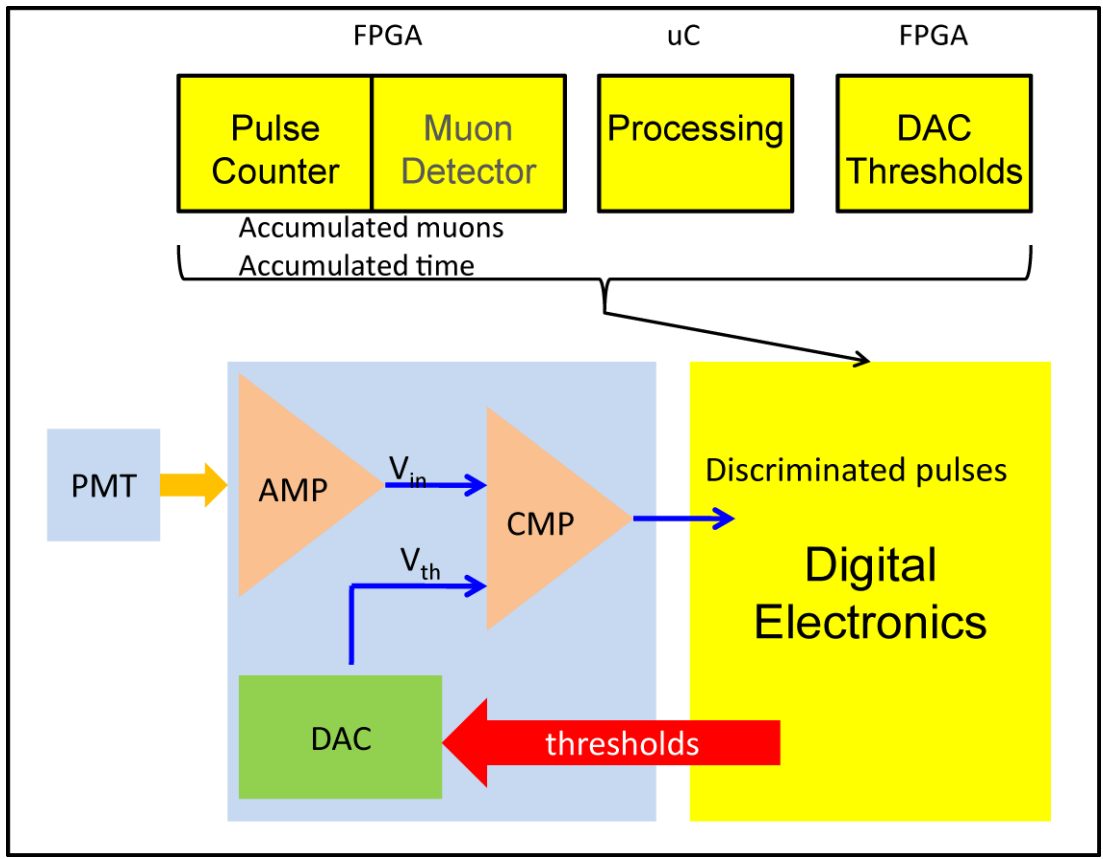

**Figure 11:** Closed loop auto-calibration process with background muons. Thresholds based on an average counting rate compensate different (and time varying) pixel gains.

## 6. Synchronization

Muon counters are used alongside surface detectors. Mass composition analysis requires correlated data from both. As surface station detectors are pre-existing and operational prior to MC installation, the synchronization scheme was tailored to work with the surface station, altering it as little as possible. The scheme can be divided in two main parts: surface to underground and underground synchronization. The former deals with synchronization of the SD to the MC surface and underground electronics plus CDAS wireless communication, while the latter is responsible for keeping the link between T1 acquired data and T3 underground data request.

**6.1 Surface station to muon counter synchronization**

MC events are synchronized to SD events through the T1 bin, and identified with timestamps. The LTS timestamp is used by underground electronics while the GPS timestamp is used by the surface electronics. The LS FPGA internal T1 bin marking pulse is sent to a synchronization module[19]. Upon receipt of an SD T1 trigger, a serial stream with the underground trigger pulse plus the 24 bit LTS is sent to a free connector pin. After electrical isolation, the 3.3 V signal is transformed to four[20] LVDS lines (one for each underground module). One dedicated category 5e twisted cable pair transmits the T1 stream to underground modules. The 10 MHz bin rate results in a total stream span of 2.9 µs. This time is fast enough to fit within a new trigger inhibition window but slow enough to avoid signal deterioration in the 25 m long cables under LVDS standard. A copy of the LTS is latched inside the LS FPGA. On event ending, a hardware interrupt forces the LS microcontroller to send the GPS timestamp to the FPGA. GTS is latched next to LTS. Under an SPI protocol, the timestamp pair is sent to the nearby surface microcontroller board through LVDS signalling. A table inside the microcontroller stores the last 2048 timestamp pairs in a circular buffer as the underground storage. When the T3 radio stream arrives with an event's data request, the table is searched for a GTS match[21]. When a match is found, the paired LTS is embedded in the event's data request to the underground electronics and is broadcasted through the Ethernet link.

**6.2 Muon Counter synchronization**

Underground electronics receives the T1 serial stream dedicated line, stripping the triggering pulse and the LTS. The first starts an event acquisition while the last tags the event. Even though signal-propagation delay in the cables and electronics is significant, the jitter is low enough to be neglected. Delay is compensated with an FPGA dedicated register that subtracts it from the total post-T1 event acquisition time. As a result, SD and MC T1 event bin times are matched within a one bin uncertainty, i.e. 25 ns. On finishing event acquisition, the LTS has already been received and linked and data are copied from the fast FPGA circular buffer to the external RAM memory bank for temporary storage. A 2048 24 bit-wide table stores the last timestamps in a ring buffer. The eleven highest external memory address bits are the same as the associated LTS position in the table, guaranteeing the link between data and timestamp. After a T3 reception by the control board, the T3 associated LTS is sent to the FPGA. The timestamp table is then searched against the received LTS. When the LTS is found in the table, its position plus an additional twelve least significant bits (LSB) zeroes, points to the beginning of associated event's data in the external memory. Afterwards, data are read into an FPGA buffer accessible from the microcontroller. An acknowledge bit, alongside a T3 found or not found, indicates to the microcontroller that it can send requested data to surface. The surface radio receives the data through the Ethernet link, transmitting it to CDAS via the Coihueco concentrator tower.

---

[19] Since nested events are not allowed after a T1 occurrence, a new trigger cannot take place before the current event ending.
[20] Eight in twin setups.
[21] GTS is formed from received data consisting of an absolute time plus an offset and a window.

## 7. Hardware validation

After initial simulations, different hardware tests were carried out. Analog and digital parts were thoroughly tested in the laboratory. Although laboratory tests are indispensable to ensure quality, there is no tougher test than the whole system stably working in the field.

### 7.1 Analog hardware test

A 500 MHz bandwidth (1.25 GHz sample rate) arbitrary wave generator[22] was used to reproduce previously acquired PMT signals with a 1 GHz, (10 Gs/s acquisition rate) oscilloscope[23]. In Figure 12, four different pulse levels were tested. The lower trace shows input pulses while the upper trace shows the discriminated (with a 10 mV threshold) signal with 3 V amplitude[24].
All injected pulses were detected.

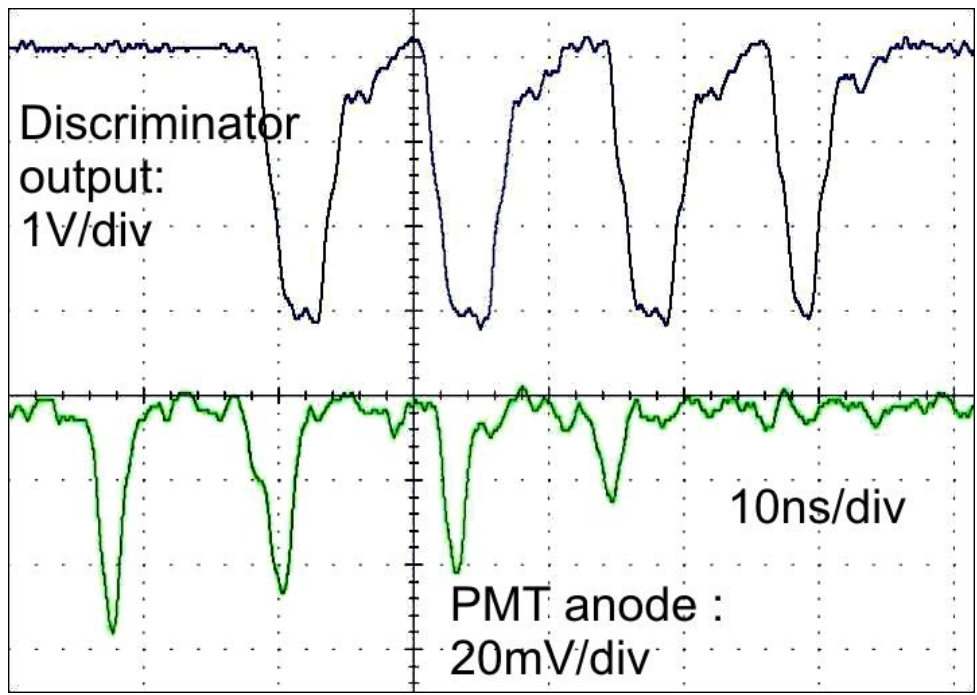

**Figure 12:** Discriminator output for four PMT signals with different amplitudes. The discriminator output of the narrowest (rightmost) pulse full width at half maximum (FWHM) is around 3 ns.

### 7.2 Digital hardware test

A 320 MHz, 64 bit-wide pattern generator was built on the digital board FPGA and connected to the 64 incoming lines. To ensure that a retrieved T3 event corresponds uniquely to the acquired event at T1 assertion, each pattern was made unique by adding a 29 bit timestamp. Moreover, 6 bits were used to identify the individual channels. In Figure 13, the first four channel signatures can be observed on the right side[25]. Each of the three consecutive runs the MSBs of the associated timestamps shows an increase as expected, and the timestamps differences correspond to T1 assertion times.

---

[22] Agilent N8242A.
[23] Tektronix DPO 7104.
[24] Delay between traces is mostly due to the discriminator delay time.

All recovered patterns were identical to the corresponding generated patterns, channel identification bits corresponded to each channel number and timestamps matched the associated T1.

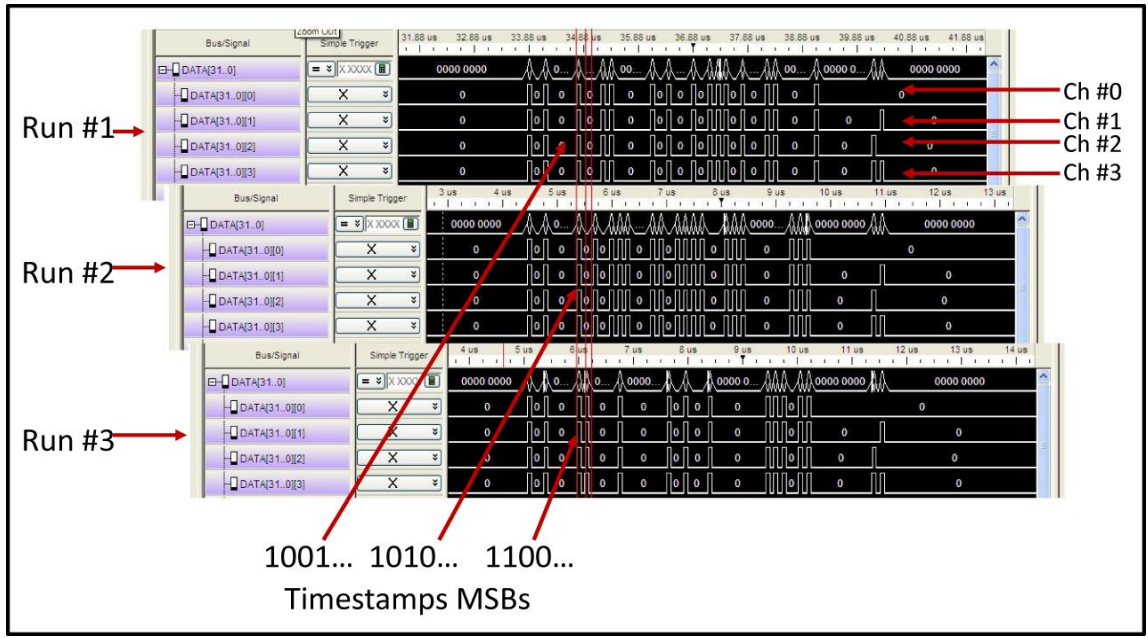

**Figure 13:** Hardware test of the digital electronics with a 64 channel 320 MHz update rate, pattern generator and logic analyzer[26]. Each pattern is made unique through a channel identification number and a timestamp.

**7.3 Data from muon counters installed at the Pierre Auger Observatory**

Figure 14 shows raw data from one counter. Each line number on the left shows a consecutive 3.125 ns sampled bin[27]. The 64 bits on the right show signal level from each of the channels; '1' represents an active state while a '0' represents an inactive state. The offline trigger strategy counts a muon when it has at least a '1', 'X', '1' pattern [17], i.e. the first sample above the discriminator threshold, the second can be at both possible states and the third is above the discriminator threshold. This ensures that only a trace with at least two photoelectrons will be counted as a muon. Dark rate and crosstalk are mostly a single photoelectron phenomenon. Different pulse amplitudes and widths of the amplified photoelectrons (due to PMT SPE distribution [18], finite amplifier bandwidth and other factors) produce different pulse lengths at the discriminator output. A pulse little wider than 3.125 ns could be read as two consecutive active samples depending on the sample timing with respect to the asynchronous analog SPE pulse. The three bin requirement eliminates this possibility. A trace is considered to be left by a single muon when its length is up to 25 ns. (Line 655 '1' falls within the 25 ns muon trace length, thus is considered to be part of M4). Dashed lines were placed where inactive samples were removed. Circles were placed on traces meeting the offline trigger criterion. On the right, detected muons were counted as M1, M2, etc. As an example, M1 spanning over bins 565 to 567 passes the offline trigger criterion while the 607 single bin does not. In this event, seven muons were counted within a 3.125 ns * (842 - 565 + 1) = 868.75 ns period.

---

[25] 64 bit recovered words are shown as two consecutive 32 bit words.
[26] Agilent 16823A, 102 channels, 1 GHz sample rate.
[27] For clarity, bins without a single active bit are omitted.

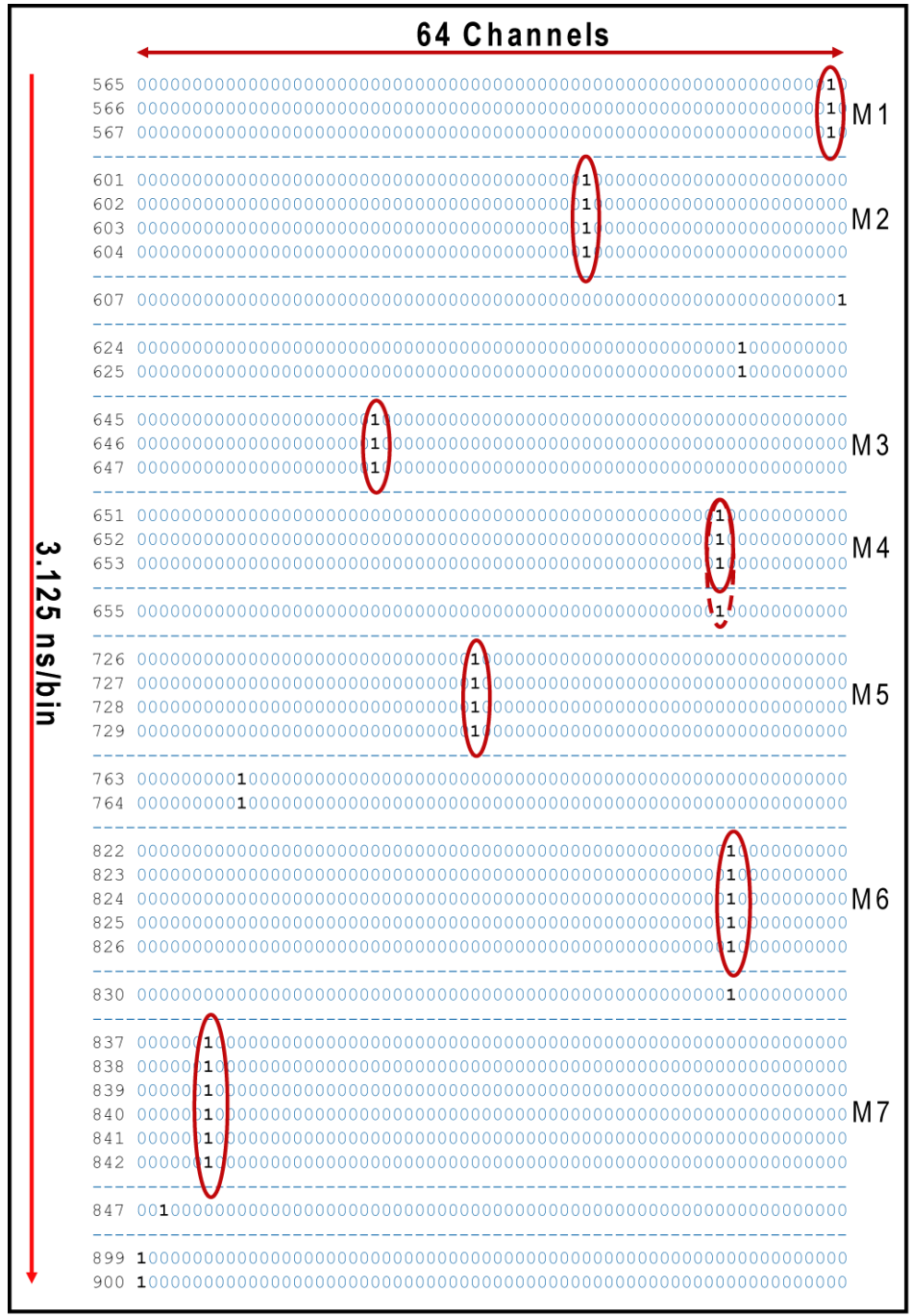

**Figure 14:** Raw data of one module from a real recovered event. Each (blue) line represents a muon counter sample. Clear muon traces (where a trace has at least a '1', 'X', '1') are circled in red.

The Local Station T1 can be triggered by different conditions. T1 Threshold is reached when the three PMT's anode signals in a Surface Station are higher than $1.75 I_{VEM}^{peak}$ [28][6] during the same 25 ns bin. T1 Time over Threshold (ToT) occurs when two PMTs anodes (out of three) are over $0.2 I_{VEM}^{peak}$ for at least 13 bins in a 120 bin sliding window. Consequently, while T1 Threshold takes a single bin (i.e. 25 ns) the T1 ToT takes at least 13 bins (i.e. 325 ns) in order to meet trigger conditions. From the scintillator module's point of view, T1 ToT events will develop at least 300 ns earlier than T1 Threshold ones.

Figure 15 shows the number of accumulated active pixels for T1-Threshold and T1-ToT trigger distributions (registered when a pixel is at '1' during a 3.125 ns bin sampling period), as a function of time (i.e. bin number) for a Unitary Cell MC[29] during a three month period[30]. Following the removal of the baseline offset and normalization, the T1-Thr (threshold) and the T1-ToT curves were integrated. 10% and 90% distribution accumulations were calculated. Table 1 shows T10 and T90 indicators corresponding to the bin number with the accumulated 10% and 90% of the total active pixel for each threshold. The peak pixel shows the pixel corresponding to the maximum of the distribution. While the T1 time bin is arbitrary, the MC stores the first sampled bin after the T1 trigger issue, regardless of the T1 type. As expected, since the T1 Threshold is triggered by a single bin in contrast to the at least 13 bins required by the T1 ToT, and its energy threshold is higher than the required by the T1 ToT [6], the T1 threshold distribution width is narrower than the T1 ToT distribution. (44 * 3.125 ns / bin = 138 ns width for T1 threshold; 185 bin * 3.125 ns / bin = 578 ns width for T1 ToT). Accordingly, since the T1 ToT is generated at least 300 ns after the T1 threshold, ToT data seen from the module T1 bin, are recorded earlier than the T1 threshold complying with the two different T1 trigger algorithms. (142 bin * 3.125 ns / bin = 444 ns).

---

[28] $I_{VEM}^{peak}$ is the peak current pulse distribution in a Cherenkov surface detector due to the light produced by the background muons. This signal is proportional to a vertical muon passing through the detector.
[29] Corresponding to SD position 688.
[30] 18/03/2003 to 19/06/2013.

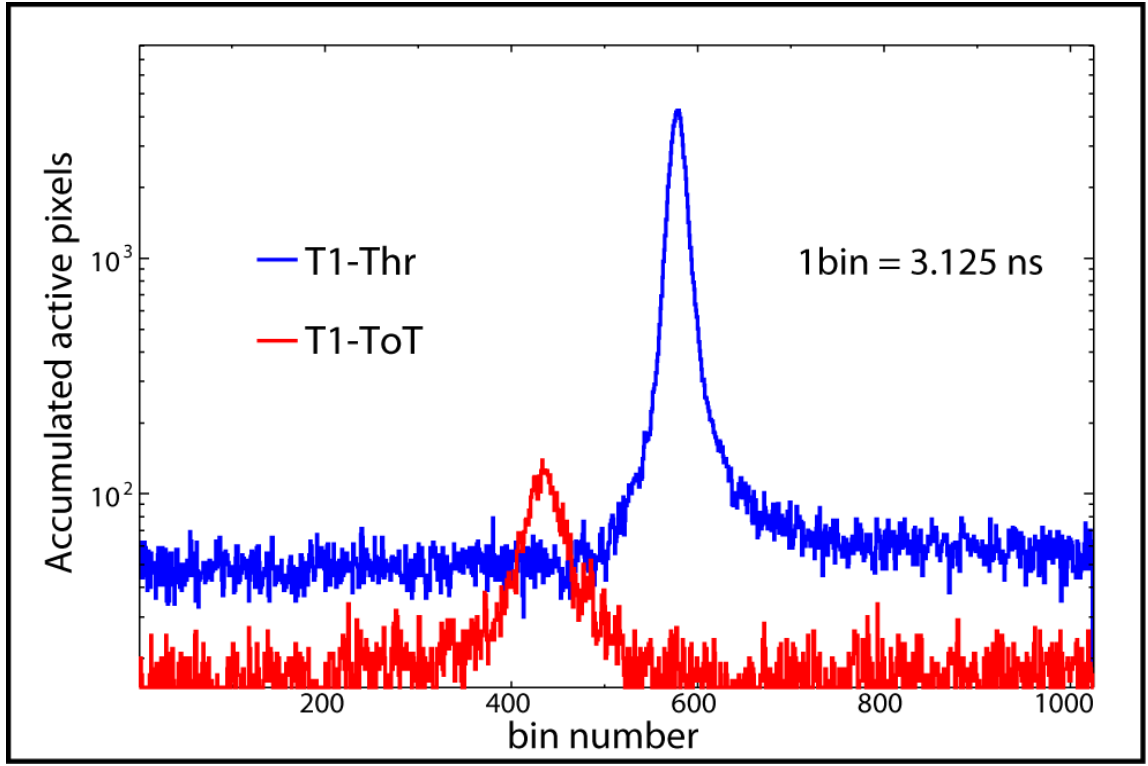

**Figure 15:** Threshold and Time over Threshold accumulated active pixel distributions, registered when a pixel is at '1' during a 3.125 ns bin sampling period, in Toune module #1 for a three month period.

| Trigger/parameter | T10 | T90 | T90-T10 | Peak position |
|---|---|---|---|---|
| **T1-Threshold** | 566 | 610 | 44 | 578 |
| **T1-ToT** | 364 | 549 | 185 | 436 |
| **ΔPeak** | | | | 142 |

**Table 1:** Fig. 13 10% and 90% accumulated active pixel distributions and peaks in 3.125 ns bin units. Baseline offset was removed and distributions normalized.

## 8. Conclusions

The AMIGA direct muon counter's digital electronics at 320 MHz (3.125 ns between consecutive samples) was described in the paper. Currently, 14 modules are successfully operating in the field.

Electronics laboratory and field tests in the Observatory were consistent with expected results. Discriminators properly convert analog PMT signals to a continuous digital stream as seen in Figure 12. Digital electronics correctly acquires time-stamped generated patterns as shown in Figure 13. Raw data show clear traces (circled '1' strings) as described in Figure 14, confirming the muon-data acquisition. The accumulated active pixel distributions, shown in Figure 15, are compatible with expected distributions and trigger timing for raw data. During the last 6 months[31], 98% of the T3 requests were responded to by the muon counters in the Observatory.

The muon counter electronics has proved to be robust and stable, working under a harsh environment in the open field, far from laboratory facilities.

---

[31] 18/03/2013 to 18/09/2013.